# Digital ‘nudges’ to increase childhood vaccination compliance: Evidence from Pakistan

Shehryar Munir[1], Farah Said[2], Umar Taj[3] and Maida Zafar[4]

**Abstract**

Pakistan has one of the lowest rates of routine childhood immunization worldwide, with only two-thirds of infants 2 years or younger being fully immunized (Pakistan Demographic and Health Survey 2019). Government-led, routine information campaigns have been disrupted over the last few years due to the on-going COVID-19 pandemic. We use data from a mobile-based campaign that involved sending out short audio dramas emphasizing the importance of vaccines and parental responsibilities in Quetta, Pakistan. Five out of eleven areas designated by the provincial government were randomly selected to receive the audio calls with a lag of 3 months and form the comparison group in our analysis. We conduct a difference-in-difference analysis on data collected by the provincial Department of Health in the 3-month study and find a significant 30% increase over the comparison mean in the number of fully vaccinated children in campaign areas on average. We find evidence that suggests vaccination increased in UCs where vaccination centers were within a short 30-minute travel distance, and that the campaign was successful in changing perceptions about vaccination and reliable sources of advice. Results highlight the need for careful design and targeting of similar soft behavioral change campaigns, catering to the constraints and abilities of the context.

[1] Viamo Inc.
[2] Lahore University of Management Sciences
[3] University of Warwick
[4] Viamo Inc.

1. Introduction

Early childhood vaccination is one of the most cost effective and efficient ways of reducing infant mortality and disability. According to estimates, nearly half of the deaths among children under five years of age worldwide are from infectious diseases that can be prevented through vaccination (WHO 2015; UNICEF 2015). Though investments in public health interventions have seen an impressive increase, nearly 20 million children do not receive routine immunization worldwide (UNICEF and WHO 2019). In Pakistan, only two-thirds of the children aged 12 – 23 months old are fully immunized (PDHS 2019). In Pakistan, and around the world, COVID-19 has disrupted routine immunization campaigns. Frontline workers, technicians and staff have been overstretched by testing, tracking, and tracing COVID-19 patients, as well as raising awareness on vaccine preventable diseases with parents of young children (UNICEF 2021).

This study focuses on a large-scale mobile-based engagement with caregivers and parents conducted by Viamo (Pvt. Ltd.) to support the efforts of the provincial health officials their efforts to promote routine vaccination for infants. The engagement was conducted in 11 Union Councils (UCs) in the district of Quetta, Pakistan, where only about one in every two children aged 2 or less are fully vaccinated. The campaign, designed after extensive interviews with caregivers and health officials, consisted of short 2 to 3-minute audio dramas sent out twice a week for 7 weeks, via automated calls. The audio dramas stressed parental responsibility for getting children vaccinated, that routine immunization is essential for the health of their children, that peers can play a positive role in encouraging vaccine take-up and compliance; and provided information on how common childhood vaccination is in the area to utilize social norms to encourage take-up.

To investigate the potential impact of the campaign on recipient vaccination behaviour, the campaign was rolled out in 6 randomly selected UCs in November 2021, allowing the remaining 5 to form the comparison group for a period of 3 months. The campaign was eventually rolled out in all 11 UCs in February 2022.

We find suggestive evidence of increased knowledge among treated UCs after the end of the campaign. We do not see a statistically significant increase in the number of children fully vaccinated in campaign areas using administrative data, but we observe interesting heterogeneity in potential treatment effects by education levels and the average travel time to vaccination sites in a UC. Specifically, we find that the number of children who are fully vaccinated did increase relative to comparison UCs in campaign UCs where the vaccination site was nearby and decreased by approximately 6% for each minute of travel time to the vaccination site. We also find that the campaign may have been more successful in increasing vaccination numbers in UCs where the average caregiver education levels were high: the number of fully vaccinated children increased by 26% over the comparison group mean by each year of average caregiver education level.

This study contributes to three key strands of literature. First, recent experiments have highlighted how predictions of traditional models of decision-making involving a careful cost benefit analysis of investments in child health do not hold, especially in resource constrained environments. This study adds to the strand of literature that tests use of behavioral science to relax some of psychological constraints in decision making (Thaler and Sunstein, 2008; Halpern, 2015, Johnson et al, 2012; Sunstein, 2013; The World Bank, 2015, Luo et al. 2021). This literature has motivated experiments testing the role of influential ambassadors, social rewards and reminders to 'nudge' adoption of desirable health behaviours (Banerjee, 2010; Johri et al., 2015; Gibson et al., 2017; Karing, 2018; Siddiqi et al., 2020; Domek et al., 2016; Banerjee et al., 2019; Hussam et al. 2021; Judah et al. 2013; Caro-Burnett et al., 2021).

Individuals can have 'biased beliefs' (Manski, 2004; Delavande, 2014) about the benefits of routine immunization, tending to overweigh immediate costs, such as, transport costs to the vaccination site or the side effects from a vaccine; underestimating the benefits that may only accrue in the future from prevention of serious disease (Chandra et al., 2019) and responding more to costs than to benefits (Kremer et al., 2019). They may also have difficulty retaining relevant but seemingly non-urgent information, especially in contexts of stress and poverty (Mani et al. 2013). Providing actional information can change behavior, even in developing countries contexts where the understanding of routine vaccines may be low (Harvey et al., 2015; Owais et al., 2011, Thornton, 2008; Banerjee et al., 2010; Kremer et al., 2019, Bronchetti et al., 2015). On the other hand, content that is unengaging or has low informational value for its audience fails to prompt take-up of preventative health behaviors (Baheti et al., 2021). The information campaign in this study draws from the insights of these three groups of studies: the audio dramas highlight the benefits of vaccination and the health costs of missing scheduled vaccination; and emphasizing the location of a local vaccination site where the government had provided vaccines for children free of cost; and it does so in an engaging way via audio dramas to draw and retain attention. We find that the information on the nearest vaccination site is potentially effective in prompting individuals to complete their child's scheduled vaccination. In the context of a developing country specifically, where access to health facilities can be poor (Ali and Altaf, 2021), our results indicate that improving the health infrastructure can improve vaccine compliance.

Second, we contribute to evidence on the use of mobile based technologies to reach large populations at relatively low cost. Digital technology, involving the use of digital platforms, such as phone or web-based applications, audio, or text messages, are increasingly being utilized in health settings, in part due to potential cost-effective mass outreach.[5] Mobile phones are portable and often shared, allowing health care providers to reach a greater number of people at relatively low cost per recipient than making individual visits (Banerjee et al. 2020). Engagement with text messages can be dependent on recipients' literacy (Wijesundara et al. 2020). In contrast, the campaign designed in this study makes use of a combination of SMS reminders for location of vaccination sites and audio dramas to reduce the dependency on recipient's literacy.

[5] SMS reminders have been found to be a cost-efficient means of increasing vaccine take-up (Milkman et al. 2011; Milkman et al., 2021; Lee et al., 2019; Regan et al., 2017); reduced dropouts from the Pentavalent vaccine schedule in Kenya (Haji et al., 2016); and improving timely child immunization completion rates in Guatemala (Domek et al., 2016), Ethiopia (Mekonnen et al., 2021), Bangladesh (Jasim Uddin et al., 2016) and Pakistan (Kazi et al., 2018). See Henrikson et al., 2018; Atkinson et al. 2019; Nkeyejyer, et al., 2021; and Eze et al. 2021 for comprehensive reviews.

Third, we add to the scare literature on mobile-based nudges to improve vaccination rates in a developing world context. Our setting is characterized by low initial levels of vaccination despite free provision of vaccines and a documented distrust of childhood vaccinations in the country (Martinez-Bravo and Stegman (2021). Within this context, the study tests if simple and engaging communication can convince recipients to comply and complete recommended childhood vaccines. We find suggestive evidence that the campaign may be most effective with those with some level of education.

The remainder of this paper is organized as follows. Section 2 describes the communication campaign, and Section 3 summarizes the data and the estimation strategy. Section 2 describes the study setting Section 5 discusses potential impact of the communication campaign, and Section 6 concludes.

## 2. Study context

Pakistan has one of the highest mortality rates in children under 5 years of age and one of the lowest vaccination coverage rates (Umer at al., 2020). Government of Pakistan's Expanded Programme on Immunization (EI) conducts about 97% of the vaccination activity in the country, providing childhood vaccination free of cost and vaccinating an estimated 5 million children under the age of one year annually (Umer et al., 2020), yet overall vaccination rates remain low. The Ministry of Health follows WHO recommended doses and timeline of vaccination, which typically mean 5 – 6 visits to the vaccination site for a full vaccination course (see Table A1). Full vaccination rates among 12 – 23-month-old children in Pakistan average at approximately 60 % (PDHS 2018), particularly due to hesitancy in Polio vaccination (Martinez-Bravo and Stegman 2021).

There is also wide provincial variation within Pakistan. Coverage rates in Balochistan, at 28.8%, are the lowest in the country (PDHS 2019). As such, Balochistan, makes for a particularly relevant setting for understanding constraints to routine childhood immunization and if low-cost, light touch interventions can improve take-up rates. Within Balochistan, our sample is defined by policy demand. We focus on 11 UCs in Quetta and Killah Abdullah that provincial health authorities highlighted as problematic areas where the district administrations had expended substantial effort to facilitate supply of vaccines, but the demand has remained lower than other UCs.

Collectively, these 11 UCs have an estimated population of nearly 600,000 with 32,000 infants aged 2 or less: 3 hospitals, 11 BHUs and 23 EI vaccination sites. Quetta, being the more populous district and the provincial capital, has almost twice the number of hospitals and BHUs than Killah Abdullah, though the number of vaccination sites are roughly equal – 13 in Quetta to Killah Abdullah's 10. There is also variation in vaccination rates within districts and across UCs - with 3 UCs in Quetta (Baleli A, Ward 10B and Ward 11B) with generally high levels of vaccination in October 2021, and poor vaccination rates in other UCs. Figure A1 in the appendix summarizes the regional variation in population of infants that is fully immunized in October 2021.

Literacy rates stand at 63% and 37% in Quetta and Killah Abdullah, respectively, while 47% and 22% of the populations of the two districts have completed primary education. More than 96% of the population in both districts have a mobile smart phone in the household (PSLM 2019). Post-natal consultations, where matters related to the health of the mother and child and recommended routine immunization, are not common in both districts, occurring 32% and 35% in Quetta and Killah Abdullah, respectively (PSLM 2019). Median household monthly income is approximately PKR 22,000 (USD 100) in Quetta and PKR 18,000 (USD 81) in Killah Abdullah (PSLM 2019).

## 3. Intervention design

### 3.1 The Theoretical design framework

Influencing health-related decisions and behaviors, such as those that govern vaccine take-up, may involve a change in caregiver behavior in different domains. Practitioner intuition can be a useful input in implementation design. For instance, in addition to informational and access constraints; and behavioural biases discussed in Section 1, social images and fear of social norms can also be relevant considerations for people deciding on course of action (Bursztyn and Jenson, 2017) and compliance with health-related guidelines (Karing et al. 2018; Siddiqi et al. 2020). Literature also suggests endorsements from ambassadors – influential local or international personalities that communities can trust – can combat misinformation and increase acceptability (Banerjee et al., 2019; Alatas, et al. 2019). Softer measures involve improving trust in the local providers, e.g., by ensuring the gender and ethnicity of the local health worker match that of the beneficiary (Berg at al., 2019).

While behavioral theories provide an explanation for the psychological and structural processes underlying human behavior and can guide design, such theories abound. A recent review highlighted 83 theories of behavior change, with overlap constructs (Davis et al., 2015). Selecting relevant guides from a wide array of theories can be a challenging task (Atkins et al., 2017). The Theoretical Domains Framework (TDF) was a concerted effort by a team of psychologists and health service researchers to develop an integrated framework for dealing with this challenge (Michie et al. 2005; Cane et al. 2012). The TDF is a synthesis of 128 theoretical constructs (Atkins et al., 2017), clustered into 14 domains: *knowledge; skills or ability; social identity; beliefs about capabilities; optimism; beliefs about consequences; reinforcement intentions; goals; memories, attention, and decision processes; environmental context; social influences;*

*emotions and behavioral regulation.*[6] The TDF helps identify the problem, the barriers that need to be alleviated, potential drivers of behavior change, and evaluating the effectiveness of possible solutions to the stated problem (French et al. 2012).

### 3.2 Design of the communication campaign

We employ the TDF for understanding mechanisms of behavior change and designing an evidenced-based communication strategy to increase vaccine take-up in the study setting. We conduct a series of interviews with healthcare providers and surveys with caregivers to garner insights regarding behaviors and practices. We used a scorecard based on the to map behaviors, beliefs and practices of beneficiaries which informed on the messaging to be created for inducing behavioral change.

We conducted semi-structured discussions with frontline workers from a hospital, a basic health unit and from a basic health unit with a maternity ward. The main purpose was to establish who came to the hospital or basic health unit and why. The first insight, that it was usually the mother of the child who came to get their children vaccinated but that they were accompanied with other women in the family, emphasized the need to stress the role of the family, rather than an individual. Health care workers found that caregivers often misplace their children's vaccination card and had trouble remembering dates of appointments on their own. However, due to socio-cultural norms, they are also not willing to share their phone numbers for reminders. Overall, the discussions complimented insights from the literature review that emphasizing caregiver's ownership over, and responsibility for, the wellbeing of their children may nudge caregivers into action. Further, due to low trust in the provider and prevailing social norms, any communication, including reminders, should be non-intrusive.

[6] The framework has been used to develop guides for implementation design (French et al., 2012) and develop theory-based questionnaires (Hujig et al., 2014). It has been cited in more than 800 peer-reviewed publications (Atkins et al, 2017), in a wide array of interventions designed to change health behaviors, such as facilitating discussions of the HPV vaccine with patients (McSherry et al., 2012), managing acute back pain (Bussieres et al. 2012), improving hand hygiene (Dyson et al. 2011), smoking cessation counseling by dentists (Amemori et al. 2011) and midwives (Beenstock et al. 2012).

We conducted automated phone surveys with caregivers of children aged 2 or less in sample areas to capture gaps in knowledge and behaviors specified in the TDF. A total of 362 individuals completed the automated survey.[7] This data was used to fill in the TDF scorecard and highlight the domains that would be relevant in nudging vaccination compliance.

Finally, we conducted a design workshop with the research team, behavioral science experts, communication strategists and health officials to finalize messaging content. Constraints highlighted by the TDF scorecard were triangulated based on three criteria: those that caregivers had the (i) capability to change; (ii) opportunity to change; and (iii) motivation to change. This led to identification of six main themes that had the highest potential of nudging a change caregivers health related behaviors: (i) knowledge on the nearest vaccination site is often lacking; (ii) getting a child vaccinated is an essential responsibility of caregivers; (iii) though it is difficult to schedule vaccinations in a busy daily schedule, it is essential for the health of their children; (iv) peers and social recognition can encourage vaccine takeup and compliance; (v) discussion on common side-effects and that the existence of some side-effects can be reassuring for caregivers; and (vi) providing information on how common childhood vaccination is in the area to utilize social norms to encourage child vaccine takeup. We then designed messages around the six themes and translated the messages in local language and vernacular to ensure high engagement.

[7] The automated phone surveys depend on the recipient to select options on the questionnaire on their own and suffer from low completion rates. In this case, the survey was started by nearly 900 individuals but completed in full by 362 individuals.

### 3.3. Implementation of the communication campaign

The communication campaign consisted of a series of audio dramas, audio messages and text messages delivered to respondents in the treated UCs over a period of 7 weeks from November 2021. Each week, the respondents would receive two audio dramas lasting 30-120 seconds to encourage childhood vaccination, centered around themes (ii) – (vi) mentioned above. In line with the first theme identified by the TDF, the calls were complimented by a combination of audio calls and text messages reminding recipients about the location of the nearest vaccination center where essential vaccines were available free of cost for their children.

We randomly selected 6 out the 11 UCs highlighted by the provincial authorities to receive the communication campaign. [8] We stratified by district so that we have representation of both districts in the campaign and non-campaign UCs, which we will refer to as the treatment and comparison UCs, respectively. We discuss the implication of this selection on our estimation strategy in Section 4.2.

We collaborated with key mobile operators in the sample areas and used geo-tagged targeting to implement the campaign in select Union Councils (UC) of Balochistan, Pakistan. We used geolocations to customize information about the nearest vaccination site in each target UC. The geolocation technology allows messaging to be sent out to all users in target geographical areas. In addition, as recommended by Viamo's Targeted Mass Messaging (TMM) service, which studies mobile usage patterns, calls were made at the time of the day that respondents were likely to be using their phones.

## 4. Data and methodology

[8] Treated UCs include 3 in district Quetta - Ward 11-A, Ward 11-B and Kharotabad 2, and 3 in district Killa Abdullah - Daman Ashazai 2, Sarki Talari and Mehmood Abad 2.

### 4.1 Datasets

Our data primarily consists of administrative data maintained at the vaccination sites by the Ministry of Health, Balochistan. This data is collected monthly and records the number of infants aged 2 years or less in the UC who have been vaccinated at each stage and the vaccine type. We observe this UC-level data from August 2021 - January 2022, i.e., 3 months before and 3 months after the intervention was implemented in November 2022.

We also use self-reported data for supplementary analysis, and to understand if the communication campaign may have impacted caregiver beliefs and perceptions about vaccines. First, like the automated survey conducted to help design the communication campaign, we conducted another automated survey with a sample of caregivers of children aged 2 or less, in all sample UCs in February 2022, three months after the communication campaign was first implemented. The survey was sent out to all respondents with a registered mobile number in study UCs and proceeded to ask detailed information only if the respondent had a child aged 2 years or less. The respondents would receive a question on SMS, select options available on the phone to record their response, and would receive the next question on the list. Of the 16,832 who responded to at least one question of the survey, 42%, or 10,415 had children under the age of 2.[9] Of these, nearly 4500 caregivers completed at least 75% of the survey. We collect information on the number of times a child under 2 in the household has been taken for routine immunization, as well as basic demographic information of households in our sample UCs in this survey. We use this data to understand if the caregiver perceptions and knowledge were different between treatment and comparison UCs nearly three months after the communication campaign was first rolled out.

[9] Based on other automated surveys conducted by Viamo, we expected a response rate of 1% ex-ante. The response rates to the request to fill a survey on SMS was relatively higher in this study, at approximately 5%.

Second, due to covid restrictions, we conduct interviews with a small sample of respondents in the treatment and comparison on the phone. We do this both immediately before the campaign implementation and three months after. Compared to the automated survey, the phone interviews allowed us to collect more detailed information on caregiver perceptions about benefits of vaccination to the child, decision making power in the household and perceived control over life events using validated psychometric scales (e.g., Falk et al, 2018). We also interviewed the same sample of respondents in both rounds of the survey to measure the difference in change in perceptions between treatment and comparison samples. Due to high levels of attrition, we were able to successfully conduct surveys with 254 in both rounds.[10] Though attrition was unrelated to treatment status – 48% of the sample in treatment and comparison UCs attrites - we consider results from this survey to be suggestive at best. Overall, attrition is related to age of the respondent, their relation to the household head, travel time to the nearest vaccination site and their preferences for risk and procrastination. Compared to the attrited sample, respondents in the balanced phone sample are 3 years older with an average age of 34, more likely to be married to the household head, is closer to the vaccination site, is more likely to procrastinate and take risk. Table A2 in the appendix provides summary statistics of the sample that attrited and the sample that was included in the final phone survey analysis.

### 4.2 Estimation strategy

The communication campaign was implemented using Targeted Mass Messaging (TMM) service at the UC level. Six UCs stratified by district, randomly selected to receive the communication campaign, form our treatment group. The remaining five UCs form the comparison sample. We conduct a difference in difference analysis, comparing the trend in vaccination numbers in the treatment and comparison samples, before and after the implementation of the communication campaign. For our main regression, we make use of administrative monthly data on the number of full vaccinations in each UC:

$$Y_{u,d,} = Treated_{u,d} + After_t + (After*Treatment)_{u,d,t} + D_d + m_t + e_{u,t} \qquad (1)$$

where $Y_{u,d}$ represents the outcomes of interest in UC *u,* district *d* and time *t*; *Treated* is a binary variable equal to 1 if UC is one of the treated, campaign UCs; *After* is a binary variable defining if the observation is from after the campaign was implemented (months November 2021 – January 2022); *D* and *m* are district and month fixed effects; and $e_{u,t}$ are random, unobserved factors present for UC *u* at time *t*. Where individual data is available, we control for respondent control for age, education, occupation, and monthly household income. For UC level data, in addition to district and month fixed effects, we include controls for the total number of infants aged under 2 (as calculated by the Balochistan Ministry of Health). All errors are clustered at the UC and month level. The coefficient on the interaction term `After*Treatment' provides an estimate of the difference in difference effect.

[10] This low response is despite multiple attempts, requests for time sent via SMS before a call was made, and incentives in the form of PKR 300 (~USD 1.4) airtime.

a. Threats to identification:

Given our analysis relies on the DD methodology for causal inference and detecting impact of the messaging campaign, it is essential that the trends in relevant outcomes be similar across the treated and comparison sample. We can indeed observe this in the monthly vaccination data, in the period of April 2020 to October 2021, before intervention implementation. As can be observed in Figure 1, trends in the proportion of the infant population that has received the full dose of vaccines in the treated and comparison UCs are similar. The vaccination rates in both types of UCs are highest at the start of 2021, potentially due to concerted efforts by the Ministry of Health to encourage routine vaccination after the lockdowns over most of 2020. They then decrease steadily over the next year, though it appears the rate of decrease may be faster in the comparison UCs than in the treated UCs after the campaign was initiated in November 2021. This may potentially be due to the communication campaign in the treated UCs.

As an additional test, we conduct an 'event study', looking at a 6-month window of the same months in the previous year, i.e., Aug 2020 - Jan 2021. We create a hypothetical treatment in November 2020 to test if we see an uptick in footfall due to treatment and see generally insignificant results (p = 0.311 for full vaccination, with no heterogeneity by education levels or average distance to a vaccination site in the UC). Results are provided in Table A3 of the appendix.

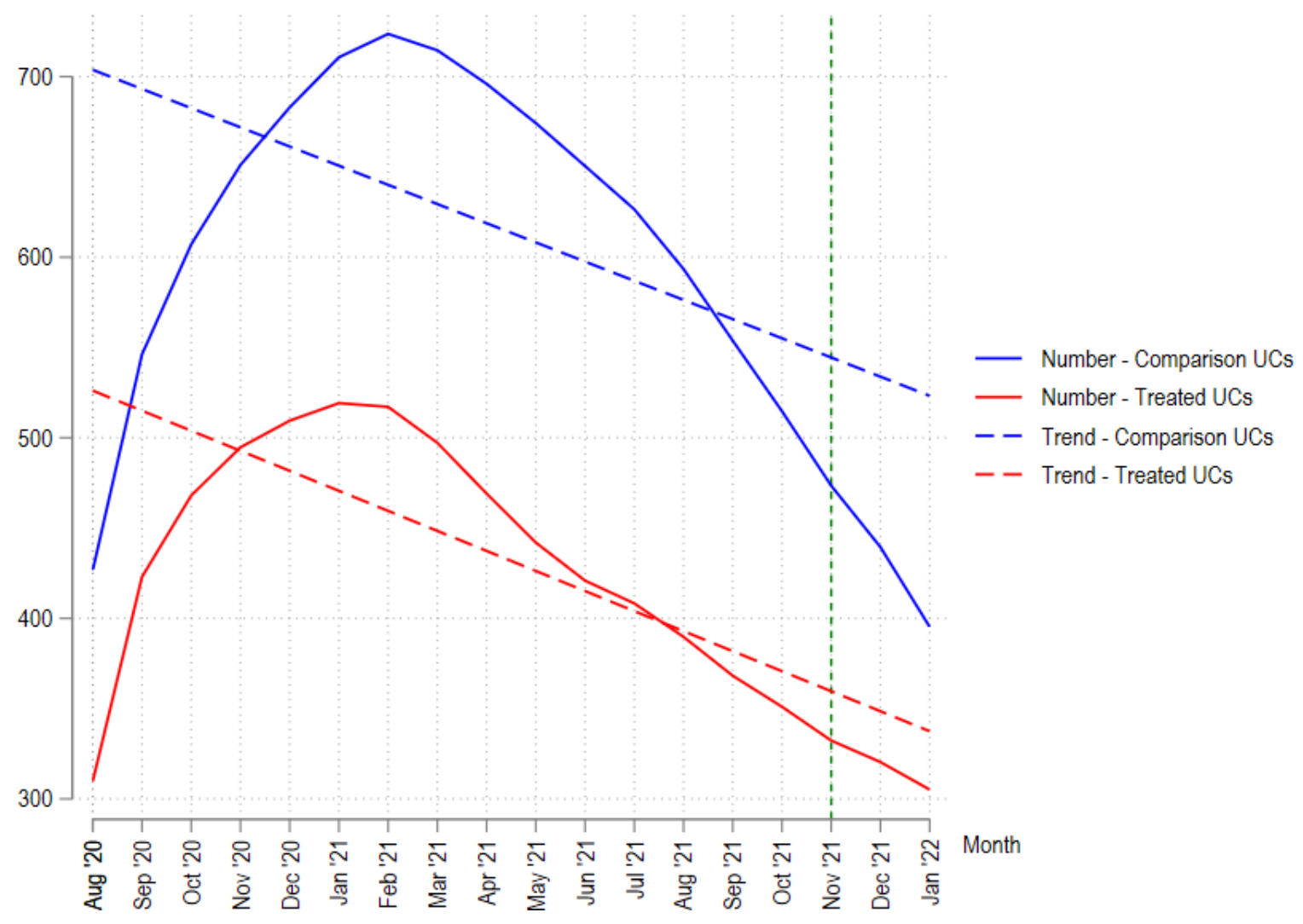

*Note: The graph plots the number of infants who were fully vaccinated between April 2020 - January 2022 in official Ministry of Health data. The green line represents the start of intervention implementation in November 2021.*

Figure 1 - Number of fully vaccinated infants in comparison and treated UCs (Aug 2020 – Jan 2022).

Note, statistical power is likely to be significantly lower when the randomization is at the UC level, rather than at the individual level. Second, literature review suggests that the impacts of a light-touch messaging information on behavior are likely to be small, which would be statistically more difficult to detect for any given sample size. Third, we observe outcomes over a relatively short period of time – 3 months after the treatment intervention is first implemented. Vaccination is not scheduled to occur continuously but is designated at specific intervals of the child's life. It is also entirely possible that change in behavior builds up over time. As a result of all these factors, we acknowledge that we may be insufficiently powered to detect a small and immediate effect.

### 4.3 Variables

#### a. Primary outcomes of interest

Our analysis will mainly focus on the number of children who have been fully vaccinated in the sample. We calculate the difference-in-differences of this outcome before and after the treatment implementation by treatment status. This outcome, measured from administrative data collected by the Ministry of Health, consists of the number of children who have completed the recommended vaccination course 3 months before and after the intervention was first implemented, i.e., between August 2021 - January 2022.

#### b. Dimensions of heterogeneity

Estimating equation (1) will provide an estimate of the potential average impact of the campaign. To capture interesting heterogeneity in how specific groups of UCs respond to the intervention, we look at two factors that are not likely to be directly impacted by the treatment but can impact the effectiveness of a soft nudge for a vaccination visit: (i) the education rates of the UC and (ii) the average distance of the government vaccination site from households in the UC. We use, as a proxy of these measures, data from the baseline automated survey that allows us to measure the average education levels in the UC and the time it takes to travel to the nearest vaccination site. On average, respondents in sample UCs have 5 years of education - or have completed primary education - and report that it usually takes them 30 minutes to reach the nearest vaccination site. Based on this information, we create indicators for UCs where:

(i) Respondents have, on average, at least completed primary education; and

(ii) Respondents, on average, report it takes them less than 30 minutes to travel to the nearest vaccination site.

c. Secondary outcomes:

Here, we will make use of automated data collected at the endline with a random subsample of residents in treatment and comparison districts, and via a phone survey conducted before and after the intervention with 254 caregivers and sample respondents. Both data can provide suggestive evidence - the former because it relies on data collected at only one point in time, and the latter because it is based on a selected sample of respondents who agreed to be interviewed twice.

However, the surveys include information that can be useful in determining if the communication campaign was successful in imparting the intended messages. We select secondary variables keeping the content of the treatment in mind. Specifically, the treatment encouraged caregivers to vaccinate their children with the support of their family and friends, to trust the health staff and family members more than religious leaders who may warn against vaccines, and to feel responsible for their child's health (rather than leaving purely to accident or fate). Finally, the communication campaign also encouraged the parents to think about the potential harms of forgoing vaccines or not complying with the recommended schedule.

Correspondingly, from the automated survey, we measure average differences in the treatment and comparison UCs on their trusted source of advice for the health of their children being:

(i) Family members, relatives or friends (peers)

(ii) Health unit staff and vaccinators

(iii) Doctors

(iv) Religious leaders

From the phone survey data, we test if the following varies over time by treatment status:

(i) The self-reported maximum number of times a child has been vaccinated.

(ii) General levels of trust in health care providers.

(iii) Individual locus of control (Pearlin and Schooler 1978), or the perception that the respondent is in control of their life outcomes; and

(iv) Perceptions about the vaccine: whether child health is above average even if the child is not fully vaccinated.

## 5. Empirical results

### a. Primary outcomes: Number fully vaccinated

We first test if the treatment campaign was successful in increasing the number of fully vaccinated infants in the treatment UCs by estimating Equation 1. Results are provided in Table 1. Two results are of note: first, as was indicated in the trends in vaccination numbers in Figure 1, we do indeed see different rates of change after the intervention in the treatment and comparison district. The DID estimate implies a significant improvement of approximately 30 children in the treatment UCs, or a meaningful increase of approximately 30% over the sample mean (column 1). The estimate is robust to the inclusion of district and month fixed effects, and for controlling for the population of infants in the district (column 2).

As explained in section 3, the intervention campaign involved 14 different waves of messaging sent by mobile operators' caregivers in treated UCs between November 2021 and January 2022. It is possible that not each call was attended. It is also likely that, of those who did attend the call, respondents did not listen to the intended message in its entirety. Treatment compliance would be imperfect for both reasons, implying that the change in footfall discussed earlier could be higher if the treatment intervention was listened to fully.

For each wave, we have data on the number of calls that connected at all, and the number of calls that connected and were listened to at least 75% through. About a fourth of the calls were attended, and two-fifths of those that connect were listened to at least 75% through, though rates vary between districts and UCs. Figure A2 summarizes both `compliance' measures in all 6 treatment UCs.[11] Unlike treatment status, compliance status is not exogenous - it is a function of individual and UC level characteristics that affect the likelihood of individuals engaging with digital campaigns. To understand if the impact of the campaign differs by compliance, we instrument compliance by treatment status. Results are provided in Table 1, columns 3 and 4, and show that for every percentage point increase in the connection and listening rates, the number of fully vaccinated children in treatment UCs increase over the change in time in comparison UCs by 1 and 2, respectively. In other words, these results suggest that, as expected, the treatment effects observed in columns 1 and 2 are stronger if residents attend or listen to the communication campaign calls.

[11]More than 300,000 respondents were targeted for the treatment campaign. As can be expected, both the number of calls attended and listened through, taper off over time and successive waves (Figure A2, panels a and b). We also find that shorter audio messages and personalized audio calls, such as those that informed respondents that a vaccine was waiting for their child at the nearest vaccine center (with location details), had better engagement.

**Table 1 - Difference-in-differences estimate of treatment campaign on the number of fully vaccinated infants in the study period**

| *Y: Total infants fully vaccinated* | (1) | (2) | (3) | (4) |
|---|---|---|---|---|
| After*Treatment | 29.267*** | 29.267** | | |
| | (7.050) | (11.282) | | |
| Connected (percent point) | | | 1.002*** | |
| | | | (0.079) | |
| Listened 75% (percent point) | | | | 2.497*** |
| | | | | (0.319) |
| District, month fe | No | Yes | Yes | Yes |
| Controls | No | Yes | Yes | Yes |
| Mean (control) | 98.967 | 98.967 | 98.967 | 98.967 |
| Observations | 66 | 66 | 66 | 66 |

te: These regressions are run on UC level administrative data covering the period 3 months before and after the TMM campaign was launched in November 2021, i.e., Aug '21 - Jan '22, for the total number of infants who receive the full course of vaccination. 'After*Treatment' is the intersection of 'Treated' defining if the UC was a treatment UC, and 'After', defining if the period is after the (placebo) treatment, and is the DID estimate.. 'Mean (control)' is the average value of the dependent variable in the control UCs over time. 'Connected (percent point)' is the percent of the study population for whom at least one of the communication calls connected. 'Listened 75% (percent point)' is the percentage of the population that listened to at least one call 75% through. Connected and listened through have been instrumented by the treatment status in column 3 and 4. Columns 2- 4 control for the population of infants aged 2 or less in the UC in Aug' 21, and district and month fixed effects, with errors clustered at the month and UC level. *** p< 0.01, ** p<0.05, * p<0.1.

b. Heterogeneity by average UC characteristics

Next, we test if the change in footfall may vary by UC characteristics. Specifically, we test if the communication campaign to ‘nudge’ footfall varies by whether the average education attainment in the UC is at least primary level, and if the average travel time to the nearest vaccination site respondents are aware of is less than 30 minutes.

Results are provided in Figure 2. In Panel (a), we see that the average impact of the treatment campaign does not vary by average education levels - the coefficients are small and precisely 0. On the other hand, when we explore heterogeneity by travel time to the nearest vaccination site, results suggest that the average impact we observed earlier may have been driven by UCs where the vaccination sites are conveniently close. In UCs where respondents report having a vaccination center within 30 minutes travel time, the treatment campaign led to an increase of approximately 70 fully vaccinated children in the study period.

a. Secondary outcomes: Differences in perceptions and preferences for advice

We now turn towards data from the automated surveys conducted at the design phase, before the treatment implementation, and in February 2022, i.e. 3 months after the implementation began. Both surveys consist of different samples of respondents in the comparison and treatment districts that were randomly selected to be sent the survey, and then responded to the questions provided. Since these are different samples, we cannot track individuals across time and analyze DID effects. However, the datasets provide us with insights on preferences the communication campaign may have changed that can underlie the average effects discussed in section 5a.

**Figure 2 - Heterogeneity in treatment effects by education levels and travel time to nearest vaccination site.**

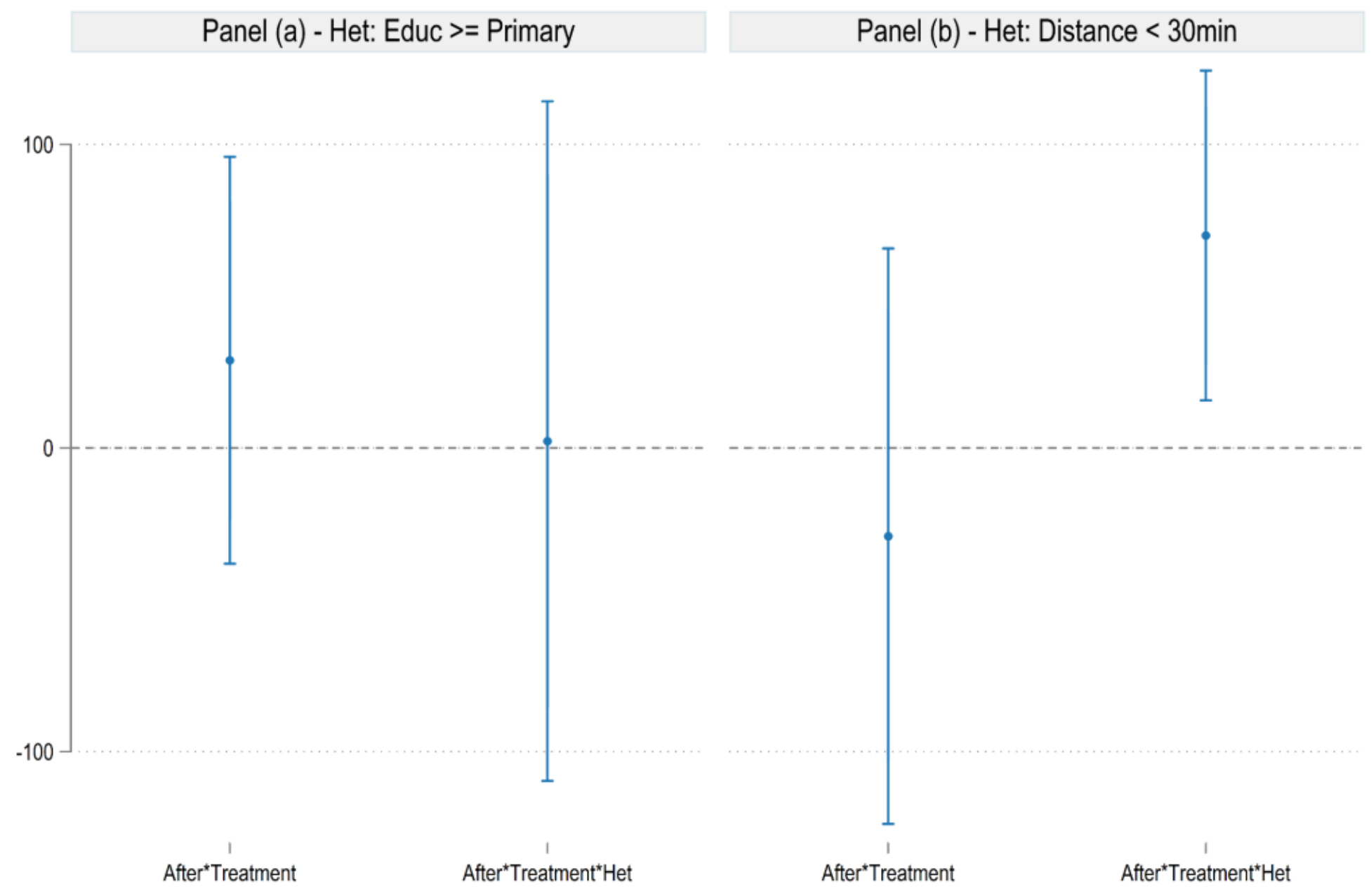

Note: The figure plots coefficient from an estimation of Equation 1 with binary indicators for dimensions of heterogeneity ('Het') and their interaction with Treatment Status and period after the implementation ('After*Treatment*Het'). In Panel (a), 'Het' is equal to a binary indicator for whether the average education level is primary or more; in Panel (b) 'Het' is equal to a binary indicator for if the average travel reported by respondents to the nearest vaccination site is 30 minutes or less. Regressions corresponding to these graphs are given in Appendix Table A4.

We test if the preferred source of advice is different across respondents in the treatment and comparison UCs, separately for each time period. In each regression, we control for individual years of education, occupation, gender and monthly household income levels and include UC level fixed effects. At baseline, we find statistically insignificant differences between treatment and comparison respondents in the likelihood of approaching family or friends, health workers, doctors or religious leaders for advice. Results are shown in Panel (a) of Appendix Table A5. However, when we compare the average preference at endline, we see that, in line with the general messages of the communication campaign summarized in Section 3, respondents in the treatment UCs are 2.3 percentage points more likely to prefer advice on child health care from family and friends, and 1.9 percentage points less likely to prefer the advice of a religious leader (Appendix Table A5, Panel b). These differences are statistically significant at the 5% level.

These results suggest that the communication campaign may have changed preferences and mindset, though we acknowledge that the results from this exercise do not present causal evidence. We have reason to believe that the campaign may have led to changes in perception. We see similar indications of the treatment campaign changing preferences and mindset from the phone survey data. As explained in section 4.1, the phone survey suffers from substantial attrition, which though unrelated to treatment, means that the final sample for whom we have data both before and after the intervention is a selected sample that does not represent the average resident of our sample. Among this selected sample, we see that though there is no increase in treatment UCs in the (self-reported) number of maximum visits any of their infants has had (Appendix Table A6, column 1),[12] treated respondents display lower levels of trust in public workers but are more likely to display higher levels of 'locus of control' or a belief that they have some degree of control over what happens in their life (Columns 2 and 3, Appendix Table A6). Combined with the automated survey data results, we take this to indicate that the intervention may have led to individuals to have greater belief in the advice of friends and family and in their own self efficacy and ability to influence life events, and less dependence on the advice of health workers and other public officials. Finally, we ask respondents if they believe a lack of vaccine will impact child health (height). Treated respondents are less likely to think that height will not be impacted. In other words, treated respondents in this sample are more likely to believe that not completing the vaccine course will lead to adverse effects. They are suggestive at best but can provide useful insights into the potential drivers of effectiveness of this, or similar, messaging campaigns.

[12] The question asked of the respondent was what is the maximum number of vaccine related visits an infant in your household has had?

## 6. Conclusion and discussion

Mobile-based communications are being used for social and behavior change campaigns around the world but little is known about its efficacy in promoting routine childhood immunization in a setting with low access and low demand of vaccination. In this study, we design and test the impact of a messaging campaign built on via multiple audio drama and audio messages over a period of 7 weeks, stressing themes around vaccination such as parental responsibility, social norms and recognition, and reminding caretakers of the location of the nearest vaccination site and their right to free vaccination.

The messaging campaign reached a high number of caretakers with relatively significant levels of geographic precision. The relatively low-cost per unit intervention has the potential to become an efficient technique to increase vaccination when complemented with public health interventions that ensure a consistent supply of vaccines. Results indicate that the soft nudge to action is likely to work when the relative cost of action is low - treatment impacts are higher when the travel time, and hence cost, of traveling to the vaccination center is low. It is also more likely to work most effectively when respondents when treatment compliance was high, i.e., when caregivers are likely to have listened to the messages in the audio drama.

We have suggestive evidence that the campaign may have changed mindset at least in the short run. Three months after the treatment campaign was first implemented, treated respondents are more likely to trust family and friends for advice, and are less likely to trust religious leaders. They are also more likely to believe that forgoing the vaccine can have adverse consequences for the health of the child.

At the policy level, results indicate that while soft nudges may convince people about the need for preventative care, they cannot operate in a vacuum. Behavior is more likely to change if it involves low time and monetary costs - e.g., if taking children for vaccination does not impose high costs on caretakers' time and financial resources. In addition, anecdotal data from vaccine centers in the study areas reveals most of the caretakers who bring their children for vaccination are women. Mobility constraints, hence, the proximity to a vaccine center, as well as support of family members may be an important channel of behavioral change.

These lessons can also be applied to increase take-up and completion of childhood vaccination courses. Similar campaigns can be particularly useful in countries such as Pakistan where misinformation is proliferated at a large scale, further exacerbating the already existing distrust surrounding immunization and vaccination. However, the effectiveness of campaigns may be limited if the intended audience has limited or inconsistent access to digital platforms, highlighting the need for careful design of information messaging campaigns to focus on the needs and abilities of the target audience.

# APPENDIX

## Additional Tables

**Table A1 – Vaccination schedule (Ministry of Health)**

| | Recommended age of the child at vaccination | Vaccination type |
|---|---|---|
| 1 | 0 months (at birth) | • OPV 0<br>• Hepatitis-B<br>• BCG |
| 2 | 6 weeks | • OPV-I<br>• Rota-I<br>• PCV-I<br>• Pentavalent-I |
| 3 | 10 weeks | • OPV-II<br>• Rota-II<br>• PCV-II<br>• Pentavalent -II |
| 4 | 14 weeks | • OPV-III<br>• IPV-I<br>• PCV-III<br>• Pentavalent -III |
| 5 | 9 months | • IPV-II<br>• Typhoid<br>• Measles-I |
| 6 | 15 months | • Measles II |

**Table A2 – Attrition in the phone survey**

| | (1) | (2) | (3) |
|---|---|---|---|
| Group: | Did not attrite | Attrited | Difference |
| | Mean/SE | Mean/SE | (1)-(2) |
| Lives in treated UC | 0.484 | 0.475 | 0.010 |
| | [0.031] | [0.019] | |
| Respondent is a male | 0.902 | 0.902 | -0.000 |
| | [0.019] | [0.011] | |
| Respondent's age | 34.102 | 31.945 | 2.157*** |
| | [0.671] | [0.384] | |
| Respondent is infant's father | 0.736 | 0.681 | 0.055 |
| | [0.028] | [0.017] | |
| Respondent is infant's mother | 0.087 | 0.083 | 0.004 |
| | [0.018] | [0.010] | |
| Respondent years of education | 8.307 | 7.868 | 0.439 |
| | [0.306] | [0.192] | |
| Respondent is self employed | 0.327 | 0.323 | 0.004 |
| | [0.029] | [0.018] | |
| Respondent is unemployed/looking for work | 0.047 | 0.070 | -0.023 |
| | [0.013] | [0.010] | |
| Number of rooms in the house | 4.694 | 5.011 | -0.317 |
| | [0.246] | [0.166] | |
| Number of household members | 13.467 | 12.956 | 0.511 |
| | [0.514] | [0.306] | |
| Monthly household income (PKR) | 46444.444 | 46952.055 | - 508.000 |
| | [2373.404] | [1518.113] | |
| HH member is a EHSAAS/BISP recipient | 0.154 | 0.140 | 0.013 |
| | [0.023] | [0.013] | |
| Respondent is household head | 0.496 | 0.452 | 0.044 |
| | [0.031] | [0.019] | |
| HH head is respondent's spouse | 0.043 | 0.015 | 0.028** |
| | [0.013] | [0.005] | |
| Respondent owns phone | 0.933 | 0.919 | 0.015 |
| | [0.016] | [0.010] | |
| Respondent uses phone multiple times a day | 0.862 | 0.874 | -0.011 |
| | [0.022] | [0.012] | |
| Child age (months, average) | 12.666 | 12.807 | -0.141 |
| | [0.395] | [0.236] | |
| Travel time to nearest EPI (hours) | 0.385 | 0.483 | -0.099** |
| | [0.029] | [0.027] | |
| Share of household decisions in which opinion is considered | 0.644 | 0.612 | 0.032 |
| | [0.021] | [0.013] | |
| Tendency to procrastinate | 8.004 | 7.068 | 0.936*** |
| | [0.194] | [0.141] | |
| Tendency to take risk | 6.180 | 5.540 | 0.640** |
| | [0.247] | [0.158] | |
| Observations | 254 | 712 | |

Note: The table reports mean and standard error of baseline characteristics reported in each row for the attrited (column 2) and non-attrited group (column 1). The value displayed in column (3) are the differences in the means across the groups. Stars indicate result from a t-test on difference on means. ***, **, and * indicate significance at the 1, 5, and 10 percent critical level. All variables have 966 observations at baseline, except measures for household income, time to EI center, procrastination and taking risk, for which we have 809, 760, 961 and 948 observations, respectively.

**Table A3 – Placebo test: DID Aug 2020 - Feb 2021**

| *Y: Total infants fully vaccinated* | (1) | (2) | (3) |
|---|---|---|---|
| After*Treatment | -26.012 | -24.447 | -10.273 |
| | (15.862) | (17.545) | (19.906) |
| After*Treatment*Primary | | 0.000 | |
| | | (0.000) | |
| After*Treatment*Distance<30min | | | 0.000 |
| | | | (0.000) |
| Observations | 66 | 66 | 66 |

Note: These regressions are run on UC level administrative data covering the period 3 months before and after a placebo or hypothetical campaign in November 2021, i.e., Aug '21 - Jan '22. Columns 1 - 3 show the results for the total number of infants who receive the full course of vaccination. `After*Treatment' is the intersection of 'Treated' defining if the UC was a treatment UC, and 'After', defining if the period is after the (placebo) treatment, and is the DID estimate. Columns 2 and 3 show heterogeneity in treatment effects by average education levels and distance (travel time) to the nearest vaccine center in a UC. `After*Treatment*Primary' provides an estimate of whether effect vary by average proportion of individuals who have at least primary education in a UC, and `After*Treatment*Distance < 30min' does the same for UCs where the average response was that the travel time to the nearest vaccination site is less than 30 minutes. All regressions control for the total number of infants aged under 2 in the UC and include district and month fixed effects, with errors clustered at the month and UC level. *** p< 0.01, ** p<0.05, * p<0.1.

**Table A4 – Heterogeneity in treatment impact (Aug 21 - Feb 22)**

| Y: Total infants fully vaccinated | (1) | (2) |
|---|---|---|
| Het | Educ >= primary | Distance < 30 min |
| After*Treatment | 28.890 | -29.067 |
| | (26.076) | (36.868) |
| Het: | 2.259 | 70.000** |
| | (43.552) | (21.129) |
| Mean (control) | 98.967 | 98.967 |
| Observations | 66 | 66 |

Note: These regressions are run on UC level administrative data covering the period 3 months before and after the TMM campaign was launched in November 2021, i.e., Aug '21 - Jan '22, for the total number of infants who receive the full course of vaccination. 'After*Treatment' is the intersection of 'Treated' defining if the UC was a treatment UC, and 'After', defining if the period is after the (placebo) treatment, and is the DID estimate.'Het' defines the dimension of heterogeneity, which is if the average UC education level is at least primary (or more) in column 1, and if the average travel time to the nearest vaccination center is 30 minutes or less in column 2. 'Mean (control)' is the average value of the dependent variable in the control UCs over time. All regressions control for the population of infants aged 2 or less in the UC in Aug' 21, and district and month fixed effects, with errors clustered at the month and UC level. *** p< 0.01, ** p<0.05, * p<0.1.

**Table A5. Automated data results for preferred source of advice at before (panel a) and after (panel b) treatment implementation**

| | (1) | (2) | (3) | (4) |
|---|---|---|---|---|
| Will take advice from: | Family/ friends | Health workers/ vaccinators | Doctors | Religious leader |
| Panel (a): Baseline | | | | |
| Resides in treatment UC | 0.011 | -0.009 | 0.003 | -0.003 |
| | (0.008) | (0.008) | (0.005) | (0.010) |
| Mean (control) | 0.580 | 0.063 | 0.111 | 0.199 |
| Observations | 4588 | 4588 | 4588 | 4588 |
| R2 | 0.016 | 0.004 | 0.008 | 0.023 |
| Panel (b) endline | | | | |
| Resides in treatment UC | 0.023** | -0.000 | 0.001 | -0.019*** |
| | (0.008) | (0.005) | (0.005) | (0.005) |
| Mean (control) | 0.506 | 0.083 | 0.111 | 0.241 |
| Observations | 3943 | 3943 | 3943 | 3943 |
| R2 | 0.014 | 0.005 | 0.009 | 0.029 |

*Note: These regressions are run using individual level data from automated surveys at endline (Jan '22). Panel (a) provides results from the baseline data, Panel (b) provides results from the endline data. 'Resides in treated UC' is a binary indicator for if the respondent lives in a treated UC. 'Mean (control)' is the average value of the dependent variable in the control UCs over time. All regressions include controls for respondent age, education, occupation, and monthly household income, with UC fixed effects and errors clustered at the UC level. *** p< 0.01, ** p<0.05, * p<0.1.*

**Table A6. Difference-in-difference estimates of potential treatment impact using data from the phone survey.**

| | (1) | (2) | (3) | (4) |
|---|---|---|---|---|
| Y: | Maximum vaccine visits | Trust in public workers | Locus of control | Vaccine not important for height |
| After*Treatment | 0.033 | -0.186*** | 0.875*** | -0.483*** |
| | (0.046) | (0.001) | (0.125) | (0.052) |
| Mean (control) | 2.784 | 0.412 | 25.895 | 5.624 |
| Observations | 175 | 254 | 254 | 137 |

Note: The coefficient on 'After*Treated' gives the difference-in-difference estimate of treatment impact using a balanced sample of data collected on the phone. Columns 1 - 4 show results for the maximum number of vaccination visits among all infants under 2 years in the family (column 1), the self reported trust in public sector workers (column 2), locus of control score of the respondent (column 3), respondent beliefs about whether they think child height can be average or above if the child receives no vaccination (column 4). 'Mean (control)' is the average value of the dependent variable in the control UCs. All regressions include UC level fixed effects, with errors clustered at the UC level. *** p< 0.01, ** p<0.05, * p<0.1.

# Additional Figures

**Figure A1. Share of infants with full vaccination in sample UCs (October 2021)**

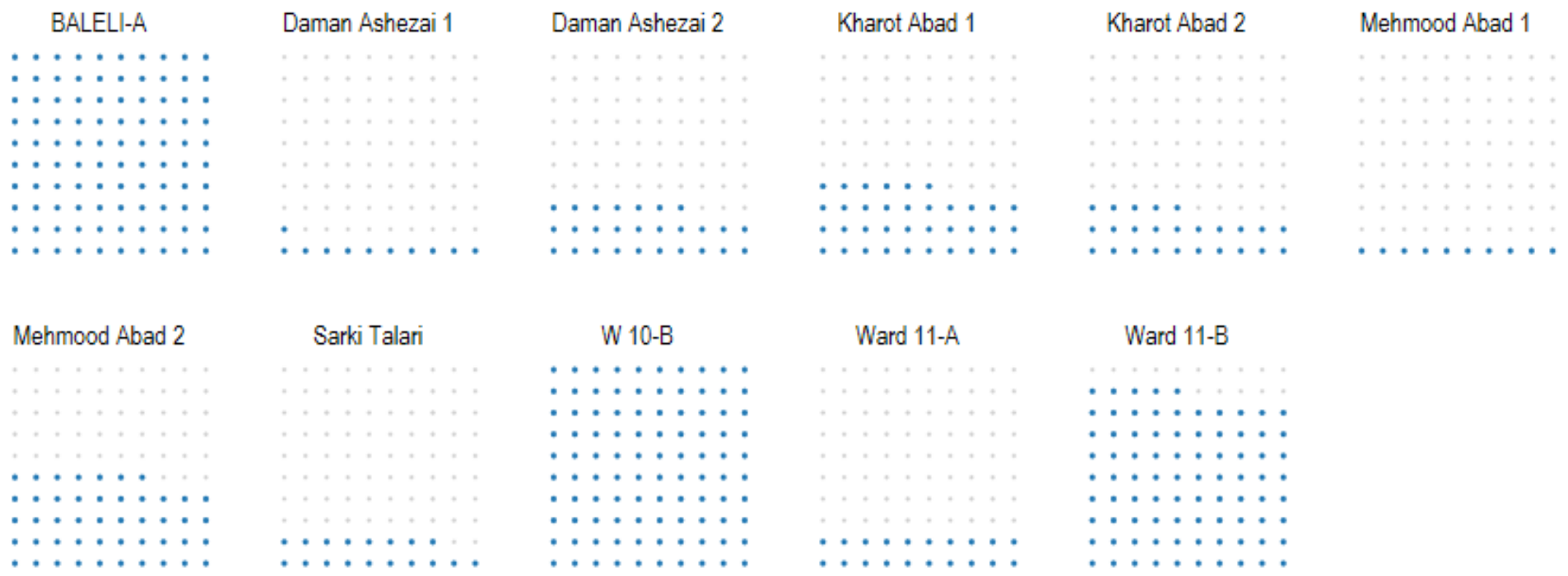

*Note: Share is % of children aged 2 years or younger in sample UCs. Each dark blue dot represents a percentage point. UCs in Quetta include Baleli A, Kharotabad 1 & 2, Ward 10-B, Ward 11-A and Ward 10-B. Source: Balochistan Ministry of Health ( % capped at 100).*

**Figure A2. Intervention TMM communication campaign connection and listening rates over time**

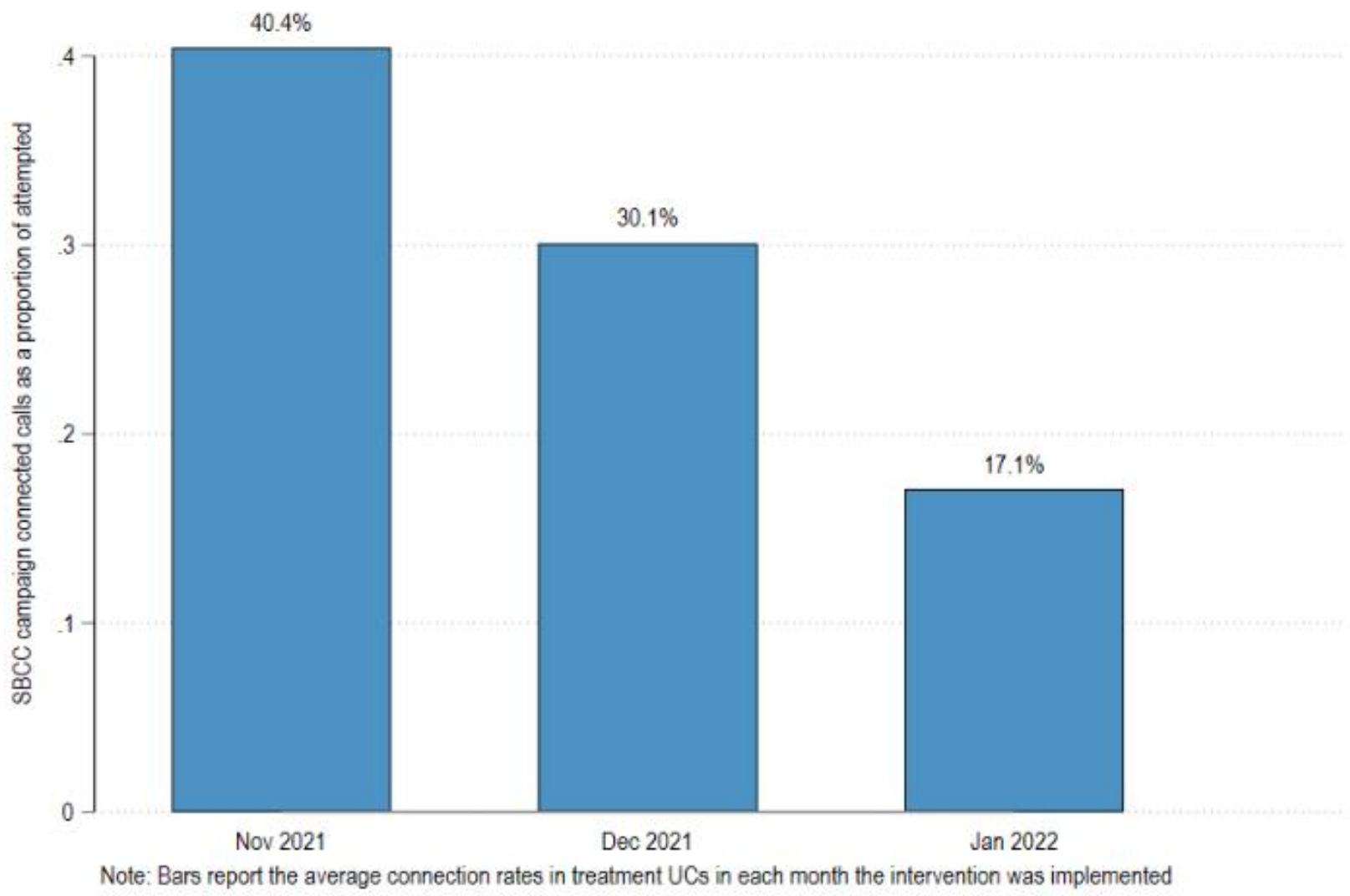

a. Calls targeted that were attended (%)

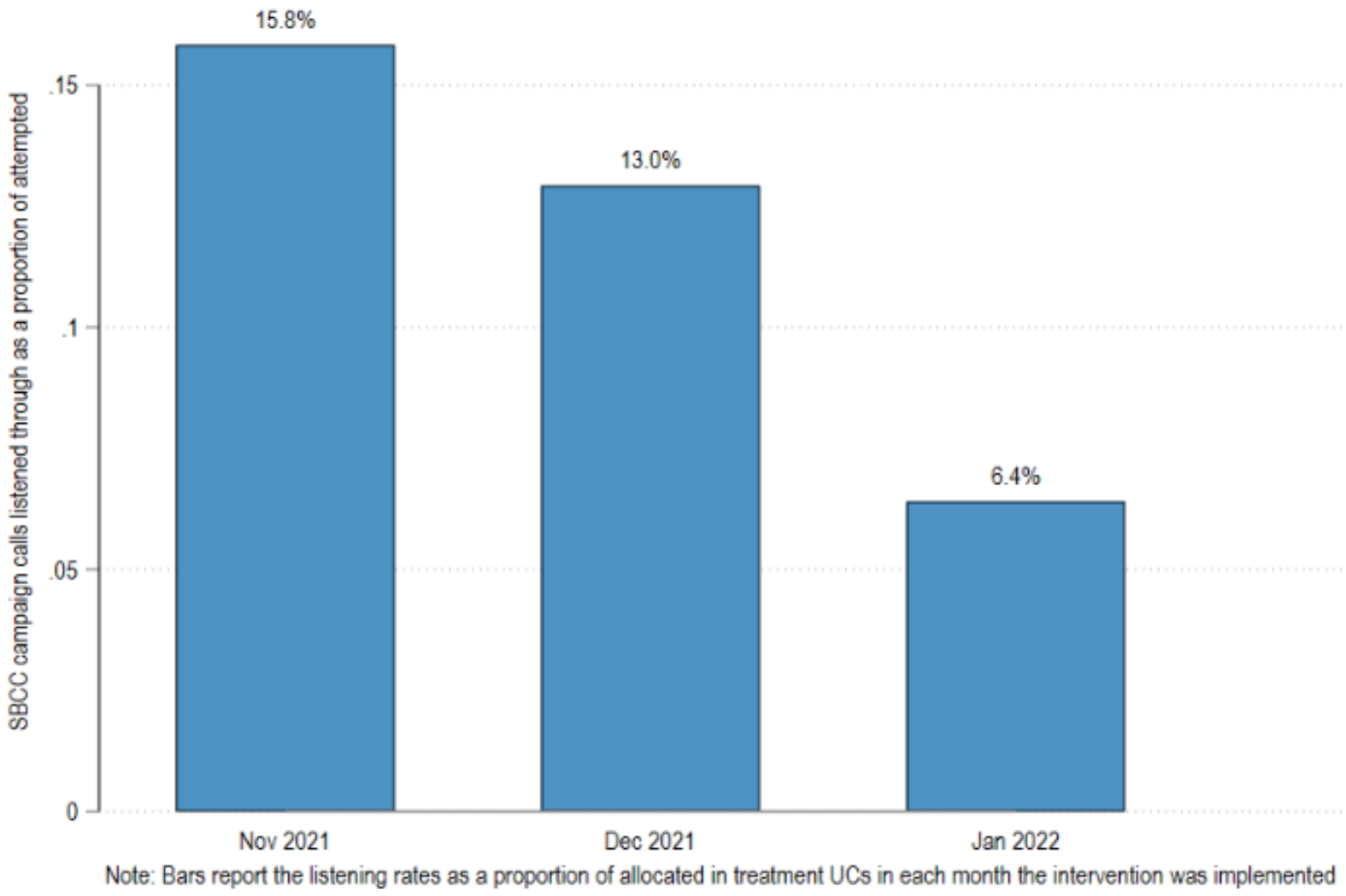

b. Calls targeted listened through (%)

*Note: The graph plots proportion of calls targeted that were attended (panel a); and proportion of targeted that were listened to at least 75% through (panel b) over the three months that the treatment intervention was implemented.*